# Single-photon nonlinearity of a semiconductor quantum dot in a cavity

D. Sanvitto, F. P. Laussy, F. Bello, D. M. Whittaker, A. M. Fox and M. S. Skolnick
Department of Physics and Astronomy, University of Sheffield, Sheffield, S3 7RH, UK

A. Tahraoui, P. W. Fry and M. Hopkinson, Department of Electronic and Electrical Engineering, University of Sheffield, Sheffield S1 3JD, UK

A single atom in a cavity is the model system of cavity quantum electrodynamics (CQED).[1] The strong coupling regime between the atom and cavity-confined photon corresponds to the reversible exchange of energy between the two modes, and underpins a wide range of CQED phenomena with applications in quantum information science, including for example as quantum logic gates and as sources of entangled states.[2,3,4] An important advance was achieved recently when strong coupling between excitons and cavity photons was reported for the first time for localized quantum dots (QDs) in micron-size solid state cavities,[5-8]. This has significance in terms of scalability and integration with other optical devices, and could lead to the emergence of 'quantum optics on a chip' technology. However the results presented so far for quantum dots are in the linear regime, corresponding to coupling to the vacuum field (vacuum Rabi splitting); they are not a true QED effect and can equally well be described by classical physics as the coupling between two oscillators.[9] In this paper, we present evidence for a purely quantum phenomenon for the QD/cavity photon system, namely the increase in splitting of the levels when the mean number of photons in the cavity is increased. This corresponds to non-linearities on the single-photon scale: the presence of a single excitation in the cavity changes the level structure, affecting the emission energies for a second photon. Such results are a first step in demonstrating the promise of quantum dots for CQED applications.

Strong coupling is achieved when the interaction energy, $\hbar g$, between a material state (atom or exciton) and a cavity mode photon is large enough to overcome the losses in the system. For QDs in semiconductor microcavities, of the type employed here, losses mainly originate from the cavity, with rate $\gamma_C$, with exciton decay rates, $\gamma_X$, typically two orders of magnitude smaller. Achieving the strong coupling regime is not straightforward, since the interaction energy of a single QD with the photon mode is typically weak, with $\hbar g$ of the order of a few tens of $\mu eV$.[10,5-7] To reduce $\gamma_C$ to this range of values requires cavity quality factors $Q = \omega_c/\gamma_c \sim 10^4$. The achievement of strong coupling, as well as high cavity Q-factors, also requires small cavity modal volumes to enhance the optical field: the condition to achieve strong coupling is proportional to the ratio $Q/\sqrt{V}$.[5]

The results presented here are obtained for an individual quantum dot embedded in a small volume, high quality factor microcavity pillar. At low pumping intensity the system is in the weak coupling regime, where irreversible decay of the excitations occurs, and the system behaves as two uncoupled states which cross in the resonance regime. This is transformed at high intensity into a doublet at resonance, characteristic of strong coupling, which transforms into a triplet as a function of exciton-cavity detuning. Good agreement is found with theoretical spectra calculated for transitions between the n = 2 and n = 1 photon states, the n = 2 state being in the strong coupling regime and n = 1 in weak coupling. The observed effects are found to be in good agreement with simulations based on the Jaynes-Cummings model[11] for the coupling between the quantum dot excitations and the cavity photons.

Our cavity is a pillar structure fabricated from a planar semiconductor microcavity grown by molecular beam epitaxy with Q-value of the planar structure in excess of 15000. It is composed of a one wavelength GaAs cavity surrounded by 27 (20) repeats of GaAs/AlAs distributed Bragg reflectors below (above) the cavity. One layer of low density, self assembled (~$5 \times 10^9 cm^{-2}$) InAs quantum dots is embedded at the centre of the cavity. The layers are fabricated into 1-3$\mu m$ size pillars by electron beam lithography and inductively coupled plasma etching. The best ratio of $Q/\sqrt{V}$ in our structures is found for elliptical pillars of dimension 2$\mu m$ x 0.8$\mu m$ (Q=8,900), as shown in Fig 1. Such elliptical pillars allow for small modal volumes, mainly confined by the small radius of the ellipse, but keeping at the same time high Qs, due to the higher content of TE polarized light of the lowest energy mode, with resulting $Q/\sqrt{V}$ superior to those in circular pillars of similar modal volume.[12] A typical pillar is shown in Fig 1: the verticality of the sidewalls and smoothness, important in retaining the high cavity Q at small volume, are noticeable.[13]

Photoluminescence experiments were carried out by exciting the pillars from the top surface with a helium-neon laser at 632nm. The energy separation between the QD and cavity mode is controlled by temperature tuning of the dot states relative to that of the cavity mode, as in refs 5-7, 13, without causing significant perturbation of the excitonic linewidth. The emission is collected from the side of the pillar at 45$^o$ to the pillar axis. In contrast to that employed by other groups, this means we collect light from both the side and the top of the pillar, affecting the observability of the features in the spectra.[14]

In Fig 2a, experimental spectra taken with an excitation power of 40$\mu W$ are shown as a function of temperature from 20 to 45K. At low temperature the cavity mode is seen (labeled C) together with an excitonic feature to higher energy (X). As the temperature is raised the exciton peak shifts to lower energy, reaching resonance at ~35K (the red trace) and then shifts further to lower energy away from resonance. This behaviour corresponds to the regime of weak coupling: at resonance the intensity of the excitonic feature is markedly enhanced, but with only one feature observed in the resonance spectrum. It is notable that away from resonance the cavity mode is populated, as in other work, probably due to background emission from wetting layer or GaAs tail states.

Very different behaviour is seen in the spectrum of Fig 2b on raising the power by a factor of 4.5 to 180 $\mu W$. In this case at low temperature the uncoupled exciton and cavity modes are again seen. However as temperature is raised and the exciton moves towards the cavity mode, a double peaked spectrum is first observed (the red trace), characteristic of strong coupling, and as temperature is increased further a 3 peaked spectrum is seen (e.g. the blue curve), before reverting to the expected uncoupled exciton and cavity modes at high temperature.

We now show that a very good account of the unusual behaviour in the near resonance regime at high power can be obtained by consideration of the coupling of a quantum harmonic oscillator, representing the cavity mode, to a two level system, a good approximation to the interband transition in a quantum dot. Without external coupling this corresponds to the Jaynes-Cummings model for atom-cavity coupling.[11,10] When the dot and cavity are exactly resonant, this predicts that the energies of the system are

$$E_\pm(n) = n\hbar\omega_C \pm \sqrt{n}\hbar g$$

For each value of $n$, the two split levels contain equal superpositions of a state with the dot empty and $n$ photons in the cavity, and one in which there is an exciton in the dot and $n$-1 cavity photons. The splitting between these dressed states increases as $\sqrt{n}\hbar g$.

External coupling is introduced phenomenologically by adding imaginary parts $\gamma_C$ and $\gamma_X$ to the cavity and dot frequencies, which broadens the Jaynes-Cummings modes by $\hbar(\gamma_X - \gamma_C)/4$. The condition to achieve strong coupling is given by $\sqrt{n}g > |\gamma_X - \gamma_C|/4$.

For the low power photoluminescence, the average number of excitations in the system $\bar{n} < 1$ so only the transitions between the $n = 1$ and unsplit $n = 0$ states are observed. This is the linear regime, where the spectrum typically consists of two lines split by $2\hbar g$ (the vacuum Rabi splitting). In the low power spectra of Fig.2(a) we do not observe this splitting on resonance, because the cavity broadening is too large for it to be resolved; the system is in the weak coupling regime. However, the splitting of the states with higher $n$ is larger, so the condition to observe strong coupling effects becomes less stringent. This explains how we are able to observe splitting on resonance in the high power spectra of Fig.2(b), where the mean number of photons is increased. In fact these spectra can be explained very well by assuming that only transitions between the n = 2 and n = 1 levels are observed. In this case, on resonance, the spectra consist of two lines split by $2\sqrt{2}\hbar g$, large enough to fulfil the requirement to observe strong coupling.

We are able to reproduce very well the behaviour in Fig 2b by solving the Jaynes-Cummings Hamiltonian as a function of the detuning between the dot transition and the cavity mode, focusing on the transitions between the $n = 2$ and $n = 1$ levels, with the n=2 level in the strong coupling regime and n = 1 in the weak coupling regime. The energy level diagram for the initial and final states of the system is shown in Fig 3a, as a function of detuning. In general, at a given detuning, $\Delta$, there are four possible transitions, whose strengths are determined by the matrix elements linking the initial and final levels, and whose linewidths are obtained from the imaginary parts of their energies. The theoretical spectra of Fig.2(c) are calculated by considering processes by which light can escape the micropillar: photons in the cavity mode leak out mostly through the top mirror or the sides of the structure, or the dot can emit directly into unconfined optical modes, most likely through the sides.[15] The relative strengths of the transitions obtained for the two types of emission differ significantly, due to the different matrix elements for the cavity and dot annihilation operators.[14] The balance between the different processes in the observed emission depends on how the spectra are measured. In particular, the multi-peak structure observed in the 45° collection angle spectra presented here is stronger than for detection through the top, implying that the 45° measurement picks up a larger contribution from the dot spectrum. The exact balance between the processes is difficult to estimate from first principles: the theoretical results we present comprise equal contributions from both.

The results of the simulations are shown in Fig 2c. All key features of Fig 2b are explained very well. The doublet feature in Fig 2b (the red trace) corresponds to resonance and arises from the transition between the strongly coupled $n = 2$ levels and the weakly coupled n=1 levels. Away from resonance, four transitions are expected of which only three have significant oscillator strength, as observed experimentally (e.g. the blue traces on Figs 2a and b).[16] Indeed over the whole range of detuning the agreement between theory and experiment is very good. The predicted variation of transition strength versus detuning is shown in Fig 3b, with the cases of on resonance (red) and close to resonance (blue) highlighted, corresponding to the spectra of Fig 2b, allowing direct visualization of the transition from doublet to triplet in Figs 2b and c, as the resonance region is crossed.

The good agreement between the theory and experiment provides good evidence that the high power spectra correspond to a situation where there are, on average, of order two photons in the system. The extra photon is not likely to arise from emission from the exciton state since this will saturate with increasing pumping power. At the same time, the cavity mode may be excited directly by emission from the weak continuum of wetting layer or GaAs tail states. Indeed the cavity feature is always visible in these and other experiments, even when there are no dots close to resonance.

An approximate estimate of the maximum number of photons which may be excited in the cavity mode is consistent with the above hypothesis. For excitation at 632nm, the absorption coefficient in GaAs is ~$3.5 \times 10^4 cm^{-1}$,[17] and an attenuation in the upper Bragg mirror of ~100 is expected. For an external incident power of 180μW, an electron-hole pair generation rate in the GaAs cavity region of ~$3 \times 10^{12}$/sec is then expected. Assuming a cavity photon lifetime of 10psec, this leads to a maximum possible photon number of ~30, although the fraction of the electron-hole pairs which subsequently lead to photon generation at the cavity mode wavelength is unknown.

Nevertheless the good agreement between experiment and theory in Figs 2 and b suggest strongly that the average photon number is >1, consistent with the maximum number obtained from the above estimates.

In conclusion, we have demonstrated nonlinear effects in a semiconductor microcavity system where the presence of an additional photon modifies strongly the emission properties of the coupled cavity/quantum-dot system. Good agreement is obtained with calculations where the system crosses from the weak to strong coupling regime as the average number of photons in the system increases. The findings provide evidence for a purely quantum effect in the optics of a quantum dot embedded in a high quality factor resonator.

Acknowledgements: This work was supported by EPSRC Grant number GR/S76076 and the EPSRC IRC for Quantum Information.

**Figure Captions**

Figure 1: Scanning electron microscope image of an elliptical pillar microcavity. The cavity structure is clearly visible showing the top and bottom DBRs and the GaAs cavity. Photoluminescence emission is collected at 45º to the cavity axis.

Figure 2: a), b) Experimental spectra of a single QD at different dot-cavity detunings obtained by increasing temperature from 20 to 45K. (a) At low power (40μW) for which only one photon is present in the cavity-dot system. The shift of the QD emission line is greater than the cavity shift with temperature, allowing negative to positive QD-cavity detunings to be spanned. The spectrum very close to resonance is indicated in red. (b) At high excitation power (180μW). The doublet trace in red corresponds to resonance in the strong coupling regime for the n = 2 state. This evolves into a triplet away from resonance (blue). (c) Theoretical spectra simulated with the strong to weak coupling transition model of Fig 3a, with g=40μeV and $\gamma_C$ = 7ps, $\gamma_X$ = 0.

Figure 3: (a) Schematic energies (left) of the dot-cavity system in the strong coupling regime for n=2 and weak coupling for n=1, as a function of detuning. The four possible transitions are indicated for a given detuning $\Delta_0$. The notation $|n, e(g)\rangle$ refers to the number of photons *n*, and either the dot populated by one exciton (e), or the ground state of the dot (g). The resulting theory spectra are shown in Figure 2c. (b) Grey scale plot of transition intensities as a function of detuning. The transitions energies are indicated by the dotted lines. Only three out of the four lines have significant strength, since the $|2,g\rangle \to |0,e\rangle$ transition is forbidden. A doublet (red, on resonance) or a triplet (blue, away from but close to resonance), are expected, as observed experimentally in Fig 2b and shown also in Fig 2c.

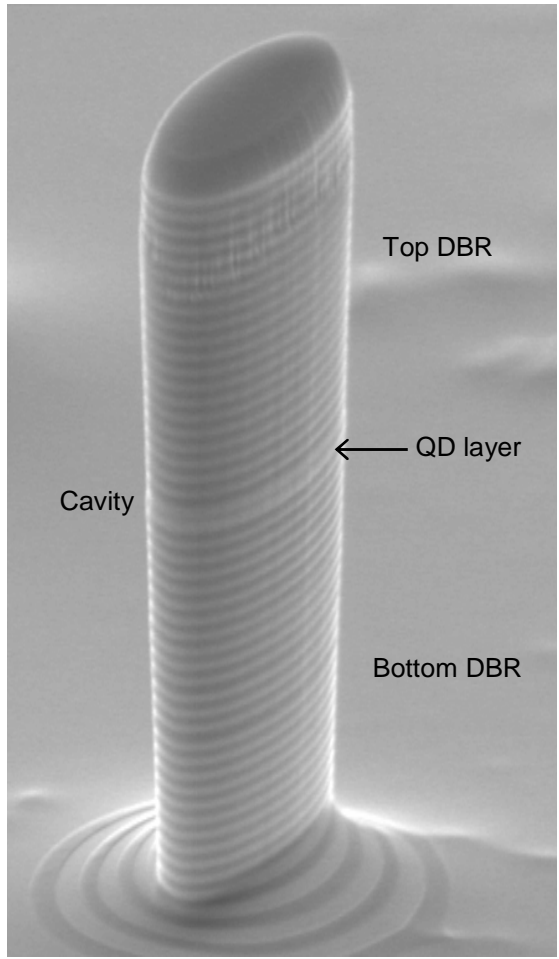

Figure 1

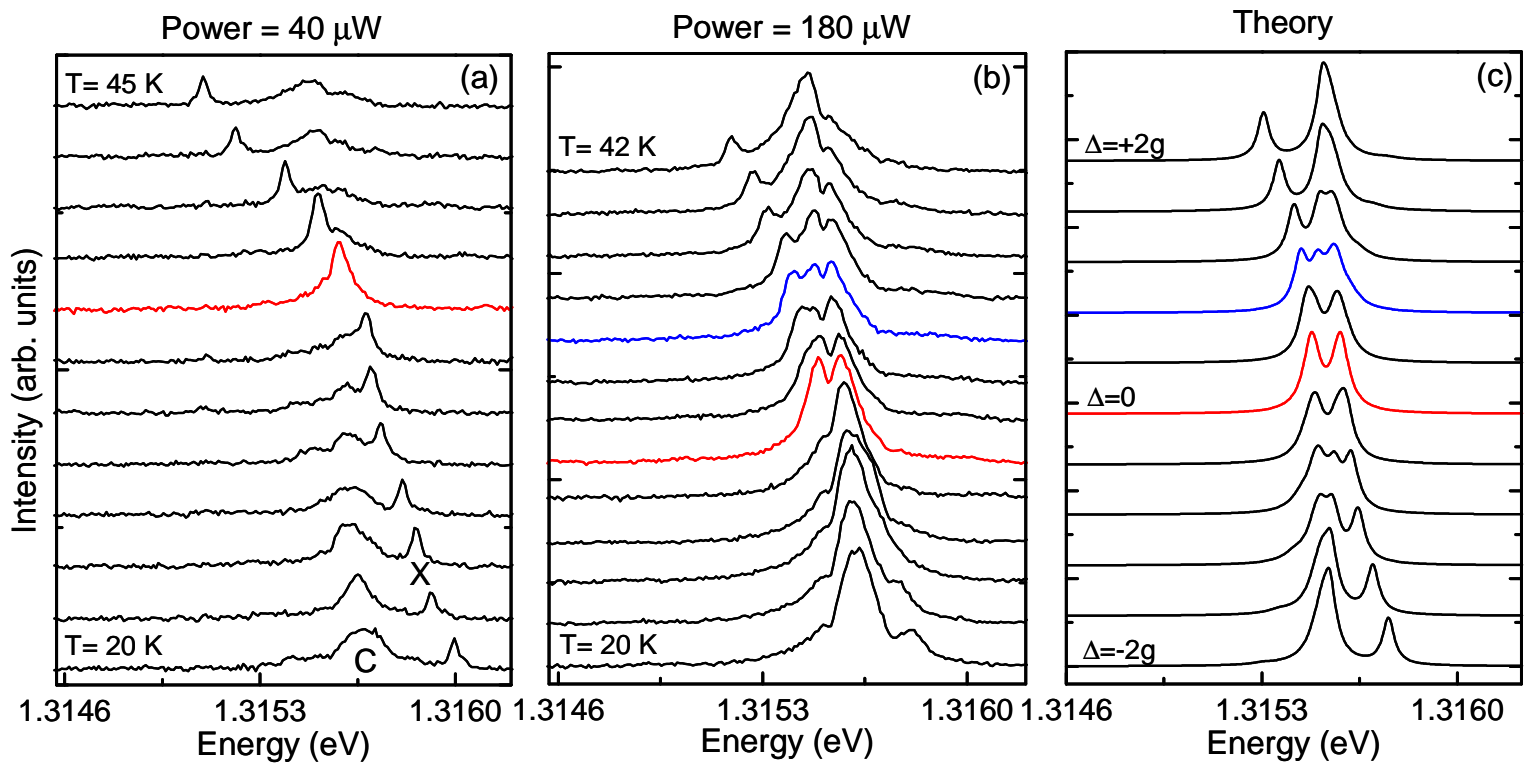

Figure 2

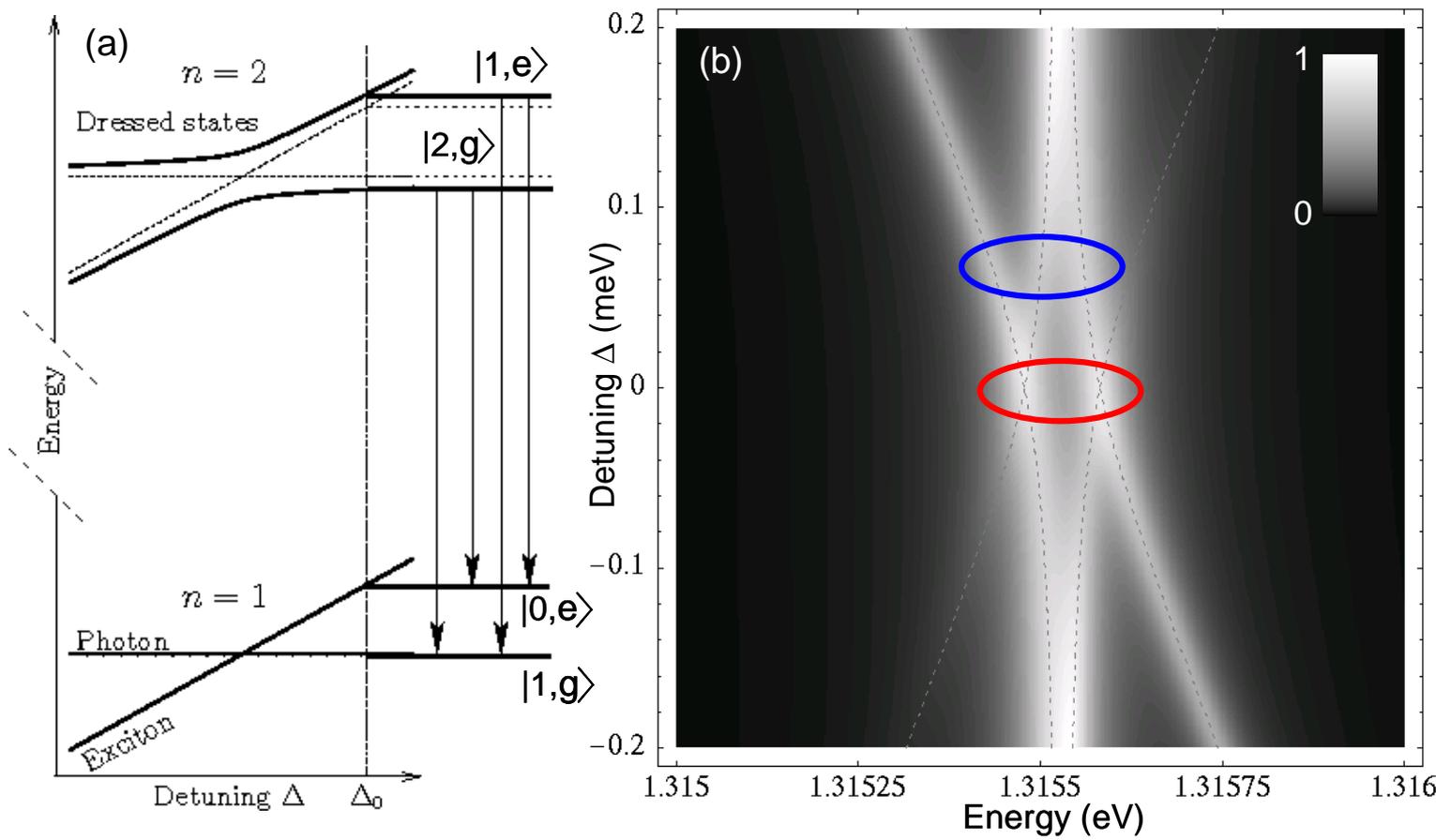

Figure 3